\documentstyle{article}
\begin{document}

\title{Using rigorous ray tracing to incorporate reflection\\
into the parabolic approximation}  
\author{Edward R.\ Floyd\\
10 Jamaica Village Road, Coronado, CA 92118-3208\\
floyd@sd.cts.com}
\date{2 September 1992}
\maketitle

\begin{abstract}
We present a parabolic approximation that incorporates reflection. 
With this approximation, there is no need to solve the parabolic
equation for a coupled pair of solutions consisting of the incident
and reflected waves.  Rather, this approximation uses a synthetic
wave whose spectral components manifest the incident and reflected
waves.
\end{abstract}

\bigskip
 
\scriptsize 
\noindent PACS NOS. 43.30D, 43.30E, 43.30G, 2.60

\normalsize

\bigskip

The (Leontovich-Fock) parabolic approximation, which approximates
the elliptic Helmholtz equation by a parabolic partial differential
equation, was originally applied to electromagnetic
propagation.$^{\ref{bib:fock}}$  Tappert and Hardin introduced the
parabolic approximation to acoustic propagation in the ocean to
account for inseparable range effects in the sound speed
profile.$^{\ref{bib:th}}$  In ocean acoustics, the parabolic
equation is a useful computational tool for tackling inseparable
indices of refraction for which the sound speed profile changes
slowly with respect to range.

One of the deficiencies of the standard parabolic approximation is
that it neglects backscatter.  Heretofore, to account for
backscatter, one solved the parabolic equation for a coupled pair
of solutions (the incident and reflected solutions).  Attempts to
account for backscatter include among others the works of Collins
et al, which uses a two-way parabolic
approximation.$^{\ref{bib:collins}}$  Herein, we present a
different approach to include backscatter.  Based on rigorous ray
tracing, we combine the incident and reflected waves into a
modulated synthetic wave that progresses in the incident direction.

Rigorous ray tracing has been developed in a generalized 
Hamilton-Jacobi representation that accounts for terms ignored by
classical ray tracing and other asymptotic methods.  It has
provided insight into propagation phenomena.  Rigorous ray tracing
has shown that the existence of a sound-speed gradient is
sufficient to induce linear (material) dispersion and angular
(geometric) dispersion even for isotropic frequency-independent
sound-speed profiles, that rays are not generally orthogonal to
wave fronts, that classical ray tracing does not predict all
caustics, and that rigorous ray tracing may be solved in closed
form whenever the corresponding wave equation may be solved in
closed form.$^{\ref{bib:floyd}}$  Its quantum mechanical analogy,
the trajectory representation, has shown how to derive the
Helmholtz equation (the stationary Schr\"{o}dinger equation) from
the generalized Hamilton-Jacobi equation.$^{\ref{bib:floyd1}}$ 
This allows us to construct the wave function or normal mode from
Hamilton's characteristic function (a generator of the motion for
the trajectory or ray path).  These normal modes can be synthetic
normal modes that contain the incident and reflected waves as
spectral components.$^{\ref{bib:floyd2}}$  We shall use such a
normal mode to develop the parabolic equation that accounts for
reflection.

Our objective in this letter is to present a parabolic equation
that accounts for reflection.  It is beyond the scope of this
letter to solve the resultant parabolic equation.  The acoustical
community is free to solve this equation by the methods of their
choice.  This work is presented in two dimensions, which is
sufficient to illustrate how to incorporate reflection into the
parabolic equation.   

We assume that the ocean to first order is a stratified medium
whose index of refraction varies with depth due to temperature and
pressure changes.  The range dependence of the index of refraction
is second order.  This index of refraction is dependent upon two
cartesian coordinates: $(x,z)$ for range and depth respectively. 
The index of refraction varies much more rapidly in the 
$z$-direction then in the $x$-direction.  We also assume that, 
for propagation of a wave train through the ocean medium, the
reflected wave is much smaller than the incident wave consistent
with the concept of backscatter.

Recently, the trajectory representation of quantum mechanics (the
quantum analogue to rigorous ray tracing) showed how the reflected
and incident waves can be combined to synthesize a wave whose front
monotonically advances in the direction of
incidence$^{\ref{bib:floyd2}}$.  The synthetic wave is given by

\begin{equation}
\begin{array}{rcl}
\underbrace{\alpha \exp[i(kx-\omega t)]}_{\mbox{incident wave}} +
\underbrace{\beta \exp[-i(kx+\omega t)]}_{\mbox{reflected wave}} &
= & [\alpha ^2+\beta ^2 + 2\alpha \beta \cos(2kx)]^{1/2} \\[.1in]
& & \ \ \times \exp \underbrace{\left\{ i\, \left[\arctan \left(
\frac{\alpha -\beta }{\alpha +\beta} \tan (kx)\right) -\omega t
\right] \right\}}_{\mbox{wave front moves in $+x$-direction}}
\end{array}
\label{eq:syntheticwave}
\end{equation}

\noindent where $\alpha$ is the amplitude of incident wave and
$\beta $ is the amplitude of reflected wave, where $|\beta |<
|\alpha |$, and where $k$ is the wavenumber and $\omega $ is the
angular frequency.  This synthetic wave is a normal mode (follows
from the superposition principle).

The synthetic wave has spatially modulation in phase and amplitude
as shown by right side of Eq.\ (\ref{eq:syntheticwave}).  For
completeness, the right side of Eq.\ (\ref{eq:syntheticwave}) was
derived from the generator of the motion for the trajectory in
Ref.\ \ref{bib:floyd2}, and the left side was subsequently
developed by analysis.  While the right side of Eq.\
(\ref{eq:syntheticwave}) was first developed from Hamilton's
characteristic function by the quantum analogy to rigorous ray
tracing,$^{\ref{bib:floyd2}}$ we subsequently learned how to do it
in a wave representation.  This is the contribution of rigorous ray
tracing that we use here.

The wave equation in two dimensions is given by

\[
\partial ^2\Psi /\partial x^2 + \partial ^2\Psi /\partial z^2 =
C^{-2}(x,z) \partial ^2 \Psi /\partial t^2
\]

\noindent The speed of sound, $C$, is isotropic and only spatially
dependent.  The wave equation is separable in time so that
$\Psi(x,z,t) = \psi (x,z) \exp(i\omega t)$  Hence, the wave
equation is reduced to the two-dimensional Helmholtz equation

\begin{equation}
\partial ^2\psi /\partial x^2 + \partial ^2\psi /\partial z^2 +
\kappa ^2(x,z) \psi = 0
\label{eq:helmholtzeq}
\end{equation}

\noindent where $\kappa (x,z) = \omega /C(x,z)$.

For reference, the standard parabolic approximation substitutes

\begin{equation}
\psi (x,z) = \theta _{\mbox{\scriptsize standard}}(x) \phi
_{\mbox{\scriptsize standard}}(x,z) = \exp (ikx) \phi
_{\mbox{\scriptsize standard}}(x,z)
\label{eq:stheta}
\end{equation}
 
\noindent into Eq.\ (\ref{eq:helmholtzeq}) to produce, after a
standard simplification, the standard parabolic equation given
by$^{\ref{bib:mcdaniel}}$

\begin{equation}
\frac{\partial ^2 \phi _{\mbox{\scriptsize standard}}}{\partial
z^2} + i2k\frac{\partial \phi _{\mbox{\scriptsize
standard}}}{\partial x} + (\kappa ^2 - k^2) \phi
_{\mbox{\scriptsize standard}}= 0,
\label{eq:spe}
\end{equation}

\noindent which does not incorporate reflection.

Let us incorporate reflection by considering 

\[
\psi (x,z) = \theta (x) \phi (x,z)
\]

\noindent where 

\[
\theta = [\alpha ^2+\beta ^2 + 2\alpha \beta \cos(2kx)]^{1/2}  \exp
\left[i\, \arctan \left( \frac{\alpha -\beta }{\alpha +\beta} \tan
(kx)\right)\right].
\]

\noindent  There is flexibility in choosing the form of $\theta $. 
Different choices of $\theta $ lead to different parabolic
equations.$^{\ref{bib:mcdaniel}}$  We have chosen a $\theta $ that
is the spatial component of the synthetic wave, Eq.\
(\ref{eq:syntheticwave}).  This $\theta $ includes reflection while
progressing in the incident direction.   In the standard parabolic
equation, the corresponding $\theta _{\mbox{\scriptsize standard}}$
in Eq.\ (\ref{eq:stheta}) is given by $\theta _{\mbox{\scriptsize
standard}} = \exp (ikx)$, which only includes the incident wave. 
Substituting $\psi = \theta \phi $ into the Helmholtz equation
leads to 

\[
\frac{\partial ^2 \phi}{\partial z^2} + \frac{\partial ^2
\phi}{\partial x^2} + 2 \frac{\partial \theta /\partial x}{\theta
}\frac{\partial \phi}{\partial x} + (\kappa ^2 - k^2) \phi = 0
\]

\noindent where $\partial ^2 \theta /\partial x^2 = k^2 \theta $ by
the superposition principle or by direct substitution.

We now examine $(\partial \theta /\partial x )\big/\theta $, which
is given by

\begin{equation}
\begin{array}{rcl}
\frac{{\displaystyle \partial \theta /\partial x}}{{\displaystyle
\theta }} & = & ik \left(\frac{\alpha ^2 + \beta ^2 - 2\alpha \beta
\cos(2kx)}{\alpha ^2 + \beta ^2 + 2\alpha \beta
\cos(2kx)}\right)^{1/2} \\[.1in]
& & \ \ \ \ \times \exp \left[ i \arctan \left(\frac{\alpha + \beta
}{\alpha - \beta} \tan(kx) \right) - i \arctan \left(\frac{\alpha -
 \beta }{\alpha + \beta} \tan(kx) \right) \right].
\end{array}
\label{eq:logdirtheta}
\end{equation}

\noindent For small reflections, $\beta \ll \alpha $, Eq.\
(\ref{eq:logdirtheta}) may be simplified to
 
\[  
\frac{\partial \theta /\partial x}{\theta } = ik [1-(2\beta /\alpha
)\cos(2kx)]\exp[i(2\beta /\alpha ) \sin(2kx)] + O[(\beta /\alpha
)^2].
\]

\noindent Now the transformed Helmholtz equation for small
reflection becomes

\[
\frac{\partial ^2 \phi}{\partial z^2} + \frac{\partial ^2
\phi}{\partial x^2} + i2k [1-(2\beta /\alpha
)\cos(2kx)]\exp[i(2\beta /\alpha ) \sin(2kx)]\frac{\partial
\phi}{\partial x} + \frac{\partial ^2 \phi}{\partial x^2} + (\kappa
^2 - k^2) \phi = O[(\beta /\alpha )^2].
\]

\noindent The critical assumption for the validity of the parabolic
assumptions is that $\phi $ is well behaved (smooth) in range so
that

\begin{equation}
\Big|\partial ^2 \phi /\partial x^2\Big| \ll \Big|2k \partial \phi
/\partial x\Big|.
\label{eq:simplification}
\end{equation}

\noindent This assumption, Eq. (\ref{eq:simplification}) is
standard for simplifying the elliptic Helmhotz equation to an
approximating parabolic equation.  The resulting parabolic wave
equation with reflection to first order in $(\beta /\alpha )$ is
given by
 
\[
\frac{\partial ^2 \phi}{\partial z^2} + i2k [1-(2\beta /\alpha
)\cos(2kx)]\exp[i(2\beta /\alpha ) \sin(2kx)]\frac{\partial
\phi}{\partial x} + (\kappa ^2 - k^2) \phi = 0
\]

\noindent or 

\begin{equation}
\frac{\partial ^2 \phi}{\partial z^2} + i2k \exp[i(2\beta /\alpha
) \exp(i2kx)]\frac{\partial \phi}{\partial x} + (\kappa ^2 - k^2)
\phi = 0.
\label{eq:pe}
\end{equation}
 
\noindent Equation (\ref{eq:pe}) is the parabolic equation that
incorporates reflection.  The difference between Eq.\ (\ref{eq:pe})
and the standard parabolic equation, Eq.\ (\ref{eq:spe}), is the
additional factor $\exp[i(2\beta /\alpha ) \exp(i2kx)]$ in the
$\partial \phi /\partial x$ term in Eq.\ (\ref{eq:pe}).  Relative
reflection as a function of the fraction $\beta /\alpha $ is
thereby incorporated to first order into $\phi (x,z)$.

In order to account for the effect of reflection, contemporary
solutions to the parabolic approximation solve the parabolic
equation for an interacting pair of solutions (incident and
reflected) or decouple the pair by
simplification.$^{\ref{bib:collins},\ref{bib:mcdaniel}}$  Herein,
we avoid the problem of coupled solutions.  Our solution to Eq.\
(\ref{eq:pe}) is a single synthetic wave that manifests both the
incident and reflected wave throughout the domain.  

The initialization of $\phi $ at some initial range, $x_i$, over
the depth column, $z$, renders the value, $\phi (x_i,z)$, over an
open boundary thereby establishing the Dirichlet boundary
conditions for a unique, stable solution for $\phi
$.$^{\ref{bib:mf}}$  This initialization process is similar to that
for the standard parabolic equation, but here we must also specify
the fraction $\beta /\alpha $.  (As a starter, one could use
Urick$^{\ref{bib:urick}}$ and the references therein to predict
$\beta /\alpha $ from reverberation and backscatter.)  Solving 
Eq.\ (\ref{eq:pe}) (which is beyond the scope of this letter) in
practice, one must not only take the usual precautions associated
with the standard parabolic approximation but also take into
account that Eq.\ (\ref{eq:pe}) is an approximation that ignores
some second-order terms of $(\beta /\alpha )$.

\bigskip

\paragraph{References}
\begin{enumerate}\itemsep -.06in
\item \label{bib:fock} V.\ A.\ Fock, {\it Electromagnetic
Diffraction and Propagation Problems} (Pergamon, New York, 1965)
Chapter 11.
\item \label{bib:th} F.\ D.\ Tappert and R.\ H.\ Hardin, ``Computer
Simulation of Long Range Ocean Acoustical Propagation Using the
Parabolic Equation Method", {\it Proceedings 8th International
Congress on Acoustics} Vol.\ II (Goldcrest, London, 1974) p.\ 452.
\item  \label{bib:collins} M.\ D.\ Collins and R.\ B.\ Evans, J.\
Acoust.\ Soc.\ Am.\ {\bf 91}, 1357 (1992); J.\ F.\ Lingevitch and
M.\ D.\ Collins, J.\ Acoust.\ Soc.\ Am.\ {\bf 104}, 783 (1999).
\item \label{bib:floyd} E.\ R.\ Floyd, J.\ Acoust.\ Soc.\ Am.\ {\bf
60}, 801 (1976); {\bf 75}, 803 (1984); {\bf 79}, 1741 (1986); {\bf
80}, 877 (1986).
\item \label{bib:floyd1} E.\ R.\ Floyd, Found.\ Phys.\ Lett.\ {\bf
9}, 489 (1996), quant-ph/9707051.
\item \label{bib:floyd2} E.\ R.\ Floyd, Phys.\ Essay {\bf 5}, 130
(1992); {\bf 7}, 135 (1994); An.\ Fond.\ L.\ de Broglie {\bf 20}
263 (1995).
\item \label{bib:mcdaniel} S.\ T.\ McDaniel, Am.\ J.\ Phys.\ {\bf
47}, 63 (1979); J.\ Acoust.\ Soc.\ Am.\ {\bf 58}, 1178 (1975).
\item \label{bib:mf} P.\ M.\ Morse and H.\ Feshbach, {\it Methods
of Theoretical Physics}, Part I, (McGraw-Hill, New York, 1953) pp.\
691, 706.
\item \label{bib:urick} R.\ J.\ Urick, {\it Principles of
Underwater Sound for Engineers} (McGraw-Hill, New York, 1967) pp.\
187--234.
\end{enumerate}

\end{document}